\documentclass[sigconf,nonacm]{acmart}
\usepackage{balance}
\usepackage[rightcaption]{sidecap}
\usepackage{xcolor}
\usepackage{latexsym}
\usepackage{tikz}
\usetikzlibrary{positioning,shapes,decorations.pathmorphing}
\usepackage{tcolorbox}

\usepackage[utf8]{inputenc} 
\usepackage{microtype}      
\usepackage{xcolor}        
\usepackage{tabularx}
\usepackage{notes-alt} 
\usepackage{multirow}
\usepackage{longtable}
\usepackage{supertabular}
\usepackage{makecell}
\usepackage{graphicx}
\usepackage{hyperref}       
\usepackage{url}
\usepackage{multirow}
\usepackage{booktabs}       
\usepackage{amsfonts}       
\usepackage{nicefrac}       

\usepackage{svg}
\usepackage{enumitem}
\usepackage{subcaption}
\usepackage{amsthm, amsmath}
\usepackage[linesnumbered,ruled]{algorithm2e}
\usepackage{xcolor}

\definecolor{blue}{RGB}{17,220,247}
\definecolor{purple}{RGB}{163,115,250}

\newcommand{\up}[1]{\tiny ($\textcolor{green}{\blacktriangle}#1\%)$}
\newcommand{\down}[1]{\tiny ($\textcolor{red}{\blacktriangledown}#1\%)$}

\usepackage{wrapfig}
\newcommand{\mpara}[1]{\medskip\noindent{\bf #1}}

\newcommand{\bert}{\textsc{BERT}}

\newcommand{\linear}{\texttt{Linear}}
\newcommand{\attention}{\texttt{Attention}}

\newcommand{\qd}{\textsc{Query2Doc}}

\newcommand{\ms}{\textsc{MS MARCO}}
\newcommand{\mshard}{\textsc{DL-Hard}}
\newcommand{\trecdl}{\textsc{TREC-DL}}

\newcommand{\ct}{\textsc{CoT}}
\newcommand{\brf}{\textit{BSF}}
\newcommand{\rrf}{\textit{R-QPP}}
\newcommand{\wrf}{\textit{W-QPPS}}

\begin{document}

\title{The Surprising Effectiveness of Rankers Trained on Expanded Queries}
\author{Abhijit Anand}
\affiliation{
\institution{L3S Research Center}
\city{Hannover}
\country{Germany}
}
\email{aanand@L3S.de}

\author{Venktesh V}
\email{v.Viswanathan-1@tudelft.nl}
\affiliation{%
  \institution{Delft University of Technology}
  \country{Netherlands}
}
\author{Vinay Setty}
\email{vsetty@acm.org}
\affiliation{%
  \institution{University of Stavanger}
  \country{Norway}
}

\author{Avishek Anand}
\affiliation{
  \institution{Delft University of Technology}
  \city{Delft}
  \country{Netherlands}
}
\email{avishek.anand@tudelft.nl}

\begin{abstract}

An important problem in text-ranking systems is handling the hard queries that form the tail end of the query distribution. The difficulty may arise due to the presence of uncommon, underspecified, or incomplete queries.
In this work, we improve the ranking performance of hard or difficult queries without compromising the performance of other queries. 
Firstly, we do LLM based query enrichment for training queries using relevant documents.
Next, a specialized ranker is fine-tuned only on the enriched hard queries instead of the original queries.
We combine the relevance scores from the specialized ranker and the base ranker, along with a query performance score estimated for each query.
Our approach departs from existing methods that usually employ a single ranker for all queries, which is biased towards easy queries, which form the majority of the query distribution. 
In our extensive experiments on the \mshard{} dataset, we find that a principled query performance based scoring method using base and specialized ranker offers a significant improvement of up to \textbf{25\%} on the passage ranking task and up to \textbf{48.4\%} on the document ranking task when compared to the baseline performance of using original queries, even outperforming SOTA model. 

\end{abstract}
\maketitle


\section{Introduction}
\label{sec:intro}

\begin{table*}[hbt!]
\begin{tabular}{lp{0.8\textwidth}}
\midrule
\textbf{Query:}  & \texttt{anthropological definition of environment}\\
\midrule
\textbf{Base Ranker (SR)} & Anthropology is the scientific study of human beings as social organisms interacting with each other in their environment, and cultural aspects [\dots] \\
\textbf{Specialized Ranker (SR)} & Environmental anthropology is a sub-specialty within the field of anthropology that takes an active role in examining the relationships between humans and their environment across space and time[\dots] \\
\midrule
\end{tabular}
\caption{Comparing top-1 document retrieved by Base Ranker and our Specialized Ranker for a hard query.}
\label{tab:intro_anecdotal}
\vspace{-5mm}
\end{table*}


Hard queries are commonly characterized by incomplete, uncommon, incorrect, domain-specific and inherently complex queries~\cite{ms_marco_hard,ms_marco_chameleons}.
Though the leaderboards of widely-used IR benchmarks such as \trecdl{} and MS MARCO indicate impressive performance gains by modern deep neural rankers, there's still significant room for improvement when addressing hard or obstinate queries \cite{ms_marco_hard,ms_marco_chameleons}. 
This paper aims to address the challenges posed by hard queries towards building effective and robust ranking models.

\mpara{The challenge of hard queries.} The presence of hard queries in training sets with large query workloads has two inherent problems.
Firstly, the number of hard queries in training sets is small, with most of the queries being relatively easier to rank.  
Secondly, the relevance factors encoded in the query-document interactions in hard queries are more nuanced and are different from those of easy queries.
As a consequence, training a single-ranking model that is agnostic to query hardness results in rankers that perform well on easy queries that dominate the training set.
However, these rankers struggle with hard or ``obstinate'' queries \cite{ms_marco_chameleons,ms_marco_hard}.
In this paper, we propose to train a specialized ranker for hard queries, in addition to the general ranker, to capture the subtle yet different query-document features induced by hard queries.
For example, as shown in Table \ref{tab:intro_anecdotal}, given a query ``anthropological definition of environment'', a base ranker retrieves a document that simply defines ``anthropology'', while a specialized ranker can retrieve a more relevant document about ``environmental anthropology''.

The presence of two ranking models -- a specialized ranker (SR) for hard queries, and a base ranker (BR) for the rest of the queries --poses two challenges :  \textit{(1) how to automatically determine which queries are hard during inference? and (2) how to train a specialized ranking model for hard queries?}

\mpara{Automated identification of hard queries}: 

We employ \textit{query performance prediction (QPP)} approaches to automatically identify hard queries~\cite{bert_qpp}.  
QPP has proved useful in determining the quality of the retrieval system on a range of queries \cite{query_difficult,query_perf,ir_need_qpp}. Prior works have also demonstrated that estimates obtained using QPP \cite{qpp_adhoc,bert_qpp,qpp_deep} are good indicators of various query characteristics, including difficulty levels. 
In our approach, queries are first ranked by both the specialized and based rankers.
This is followed by a \textit{fusion mechanism} to aggregate both scores based on a hardness estimate that is determined using QPP on the input query.
This approach of combining the merits of both rankers provides an automatic and robust way of ranking documents for any arbitrary queries.

\mpara{Context Aware Query enrichment.} Our second major challenge is training the specialized ranker for hard queries.
A ranker, trained on hard queries in their original form, does not capture the query-document features for hard queries due to the lack of context.
To help the specialized ranker easily identify the specific query-document features, our second novel contribution is to \textit{contextually expand} the hard queries.
Specifically, we hypothesize that if the hard queries are expanded or enriched with the active knowledge of the relevant documents, then a specialized ranker can easily extract relevant query-document features when training on the expanded queries.
Towards this, we propose a \textit{context-aware query enrichment approach}, to rewrite queries using LLMs. This method incorporates knowledge from the relevant documents as context during training, enhancing the rankers' ability to learn the complex relationship between query-document pairs for hard queries.

We conduct extensive experiments on \mshard{}~\cite{ms_marco_hard} (hard queries from \ms{}), and the TREC-DL~\cite{zerveas2020brown} document datasets to showcase the performance benefits of using specialized rankers.
We find that during inference, our specialized rankers are surprisingly effective for hard queries without needing any query expansions.
Our approach shows significant performance gain of \textbf{20.2\%} and \textbf{48.4\%} in nDCG@10 and RR respectively compared to baseline and outperforming the SOTA model.
We believe that this observation has an impact on the design of multiple ranking models for ranking to handle different query types.

\subsection{Related Work}
\label{sec:rel-work}

The field of IR has long explored the concept of query expansions, with both classical~\cite{arabzadeh2021matches,grbovic2015context,zamani:2017:relevance-based-word-embedding,he2016learning,rao2018learning,zerveas2020brown} and recent Large Language Model (LLM)-based methodologies~\cite{wang2023query2doc,gospodinov2023doc2query,jagerman2023query,wang2023generative, trienes2019identifying,zuo2022contextaware,zamani2020generating}. Distinct from these methods, which typically use expansion during the inference, our method uniquely applies expansion only during the training phase. 
Our technique is also different from document expansion methods~\cite{doc2query,docT5query} and document augmentation strategies~\cite{bonifacio2022inpars,dai2022promptagator,anand2023data,anand2022supervised,rudra2023depth,rudra2020distant}, which target document modification rather than query rewriting. 
Furthermore, these generative expansion approaches are context-unaware, leading to topic drift. Existing strategies to mitigate such drift~\cite{selective_query_expansion_1,selective_query_expansion_2} are again primarily implemented during inference, which is different from our approach.
\section{Methodology}

\subsection{Query Enrichment}
\label{method:car}
Given a training set, consider a set of queries $Q=\{HQ \cup Q^+\}$, consisting of a set of hard queries $HQ$ and a set of easy queries $Q^+$.
Given a hard query $hq_i \in HQ$, we generate a rewritten query $q_i^*$ that enriches the hard query to result in a more clear and concise query for a ranker.
$q_i^*$ is generated by conditioning the LLM on a prompt using the query and top most relevant document $d^+$ from a document collection $D$, known as \textit{context-aware query enrichment}. Formally, $q_i^* = \mathbb{LLM}\left(hq_i,d^+\right)$. 
Note that there are no principled methods for automatically determining $HQ$. We use heuristics such as query length~\cite{ambig_query_identity_length}, presence of uncommon (acronyms and entities with varied semantics) or incorrect terms and lack of context (under-specified) which are defined according to characteristics of obstinate queries in \mshard{} and \ms{} Chameleon datasets.


\textbf{Passage selector for documents:}
\label{method:attention_linear}
Using the full document in the prompt for query enrichment could result in topic drift, since a document may consist of multiple aspects and different topics. To mitigate this, we employ supervised passage selection techniques (\attention{}, \linear{}) as proposed in~\cite{leonhardt2021learnt}. These techniques aid in selecting the most relevant passage from a document, aligning it more closely with a given query.

\subsection{Training Ranker}
\label{method:ranker}
Next, our goal is to train a model for \textit{re-ranking} documents. We employ the pointwise loss to train the ranker. Given a query-document pair $(q^*_i, d_i)$ as input, the ranker model outputs a relevance score. The ranking task is cast as a binary classification problem, where each training instance $x_i = (q^*_i, d_i)$ is a query-document pair and $y_i \in {0, 1}$ is a relevance label. 
We train two rankers, a Specialized Ranker ($SR$) trained on hard enriched query using \textit{context-aware query enrichment} and a Base Ranker ($BR$) on all original queries. 

\subsection{Specialized Ranking of Hard Queries}
\label{method:rank_fusion}
During inference, we use both SR and BR as below:
\subsubsection{Balanced Score Fusion (\brf{}):}
\label{method:balanced_rank_fusion}
To improve ranking performance for all queries, we propose to aggregate ranked lists from the base ranker $BR$ and the 
$SR$ during inference using the CombSUM technique \cite{Fox1993CombinationOM}. More formally, for a given test query, $q_i$ and corresponding document $d_i$, 
$r_i = BR(q_i,d_i) ; r_j = SR(q_i,d_i)$.
The final relevance score is calculated as the sum of document retrieval scores: $r_i+r_j$. 
\subsubsection{Routing using Query Performance Prediction (\rrf{}):}
\label{method:qpp}
Using capabilities of $SR$ and $BR$ on specific queries should improve the ranking performance. In order to determine if a query is hard at inference time, we use a Query Performance Prediction (QPP) model inspired by BERT-QPP \cite{bert_qpp}. We train a BERT-QPP model as per the original implementation optimized for nDCG@10.
During BERT-QPP model training, given a query $q_i$ and top-k retrieved documents $D^+_k$ for the query using BM25, we fine-tune a \bert{} based cross-encoder model $\psi$  using binary cross-entropy loss as follows:
\begin{multline}
    \mathcal{L} = -\frac{1}{N}\sum_{i=1}^{N}\Bigl( M(q_i,D^+) \cdot\log \psi(q_i,D^+_k)\\
    + (1-M(q_i,D^+))\cdot\log(1-\psi(q_i,D^+_k))\Bigr)
\end{multline}

where $M$ refers to nDCG@10 score and $D^+_k$ denotes the top-k retrieved documents from all retrieved documents $D^+$.

At inference time, we score queries using a trained BERT-QPP model and use the score to decide if a query is hard or easy. Hard queries are evaluated by SR and easy queries by BR and finally combine their ranklist.

\subsubsection{Weighted QPP Scoring (\wrf{}):}
\label{method:qpp_rf}
Finally, we propose a QPP based approach which combines the two ranking scores ($BR$ and $SR$) weighted by the QPP score.
Given a query ($q_i$) at inference time, we first obtain a score which is the \textit{hardness estimate} for $q_i$ using BERT-QPP model ($\psi$) defined above given the retrieved set.
\[\mathbb{P}(hard|(q_i,D^+_k)) = \psi\left(q_i,D^+_k\right)\]
Then the hardness estimate $\psi\left(q_i,D^+_k\right)$ is used to compute final relevance score $s_i$ by interpolating the relevance scores obtained by BR or $s_{BR}\left(q_i,d_i\right)$ and SR or $s_{SR}\left(q_i,d_i\right)$.
\begin{equation}
    s_i = \psi\left(q_i,D^+_k\right) * s_{SR}\left(q_i,d_i\right) + \left(1-\psi\left(q_i,D^+_k\right)\right) * s_{BR}\left(q_i,d_i\right)
\label{eq:qpp_rf}
\end{equation}

Intuitively, the higher the QPP score, more weight will be given to the specialized model score. The resulting score $s_i$ is used to rank documents. We show in the experiments (Section \ref{subsec:qpp_rf}) that QPP based scoring improves performance on hard queries without compromising performance on other queries.

\section{Experimental Setup}
\label{experiments}
This section describes the experimental setup for three research questions --
{\textbf{RQ1}}: How effective is the Specialized Ranker ($SR$) for hard queries? 
{\textbf{RQ2}}: What is the impact of QPP in identifying hard queries?
{\textbf{RQ3}}: How effective is QPP based scoring for ranking?

\subsection{Datasets}



\subsubsection{\ms{} (Document dataset)}: We consider the document dataset from the TREC Deep Learning (DL) track (2019)~\cite{nguyen:2016:msmarco}. 
Since, LLM based query enrichment is expensive, inspired by \textit{selective query expansion} \cite{selective_query_expansion_3, selective_query_expansion_1}, we randomly sample a small set of queries that satisfy several heuristics from the \ms{} training set(described in Section~\ref{method:car}).   
For training the specialized ranker, this approach results in the selection of 1200 hard queries from \ms{} document training set, and then we rewrite these queries using our query rewrite models (Subsection \ref{models}). We evaluate our proposed model on \trecdl{}-19 and \trecdl{}-20, comprising 43 distinct queries each.
\subsubsection{\mshard{} dataset}: Contains an annotated collection of 50 hard queries collected according to several criteria defined by \citet{ms_marco_hard}. \mshard{} queries are primarily non-factoid. 




\subsection{Query Rewriting Models}
\label{models}
\subsubsection{\qd{}} 
We use a recently proposed document expansion approach, \qd{} which generates pseudo documents to aid in document ranking, as a baseline. We use the gpt-3.5-turbo model with max tokens of 128 for generation and follow the original hyperparameters used by~\citet{wang2023query2doc} for reproducibility. We further use the rewrites to train a \bert{}-base model for ranking.

\subsubsection{Chain-of-Thought (\ct{})} 
We also compare against the recent LLM based query expansion approach, which uses zero-shot chain of thought prompting to improve the query \cite{jagerman2023query}. We employ the same hyperparameters as the original paper. As above we train a ranker using these rewrites.


\subsection{Ranking Models}
\label{ranking_models}
We use the cross-attention \bert{}-base~\cite{devlin_bert_2018} architecture(12 layer) for ranking. The input length is restricted to a maximum of 512 tokens. 
The BR is trained on the original set of queries using a pointwise ranking loss objective with learning rate of 1e-5 and 3e-3 for document set respectively. SR is trained with learning rate of 4e-4 for both sets. 
For all experiments, we use nDCG@10 and RR metrics for comparison.

\begin{table}[hbt!]
    \centering
    \begin{tabular}{lll}
        \toprule

             & \multicolumn{2}{l}{\textbf{MS HARD Document}}\\
             \midrule
             \textit{Ranking Models}
             &  $\text{RR}$ & $\text{nDCG}_\text{10}$ \\
            \midrule

\multicolumn{3}{l}{\bf Baselines} \\
\bert{}-base (\textbf{BR}) & 0.444 & 0.324 \\	
\qd{}~\cite{wang2023query2doc} & 0.510\up{14.9}& 0.248\down{23.5}$^{*}$  \\
\ct{} \cite{jagerman2023query} & 0.566\up{27.5}& 0.336\up{3.7}$^{\#}$\\
BM25$^{\dag}$ & 0.368 \down{17.1} &  0.272 \down{16}  \\
BERT-MaxP(ZS)$^{\dag}$ & 0.405\down{8.8} & 0.310\down{4.3}  \\
BERT-MaxP$^{\dag}$  & 0.402\down{9.5} & 0.317\down{2.1}\\
RM3+BERT-MaxP(ZS)$^{\dag}$~\cite{lavrenko2017relevance}  & 0.415\down{6.5} & 0.314\down{3}  \\
RM3+BERT-MaxP$^{\dag}$~\cite{lavrenko2017relevance} & 0.443\down{0.2} & 0.295\down{8.9} \\
T5-MaxP(ZS)$^{\dag}$~\cite{raffel2020exploring}  & 0.367\down{17.3} &  0.327\up{1} \\
RM3+T5-MaxP(ZS)$^{\dag}$~\cite{raffel2020exploring} & 0.359\down{19.1} &  0.307\down{5.2} \\
Electra-MaxP$^{\dag}$~\cite{clark2020electra} & 0.448\up{0.9} & 0.385\up{18.9}\\
RM3+Electra-MaxP$^{\dag}$~\cite{clark2020electra} & 0.461\up{3.8} &  0.380\up{ 17.4}\\
PARADE-BERT$^{\dag}$~\cite{li2020parade} & 0.413\down{7.0} & 0.299\down{7.7}\\
RM3+PARADE-BERT$^{\dag}$~\cite{li2020parade} &  0.419\down{5.6} & 0.313\down{3.3}\\
PARADE-Electra$^{\dag}$~\cite{li2020parade}~\cite{clark2020electra} & 0.498\up{12.2} & 0.356\up{9.9}\\
RM3+PARADE-Electra$^{\dag}$~\cite{li2020parade}~\cite{clark2020electra} & 0.489\up{10.1} & 0.357\up{10.3}\\

\midrule
\multicolumn{3}{l}{\bf Our approach WITH query rewriting during inference} \\
\midrule
Specialised Ranker(SR) & 0.345\down{22.4}& 0.257\down{20.6}\\
Balanced Score Fusion(\brf{}) & 0.568\up{28}$^{*}$ & 0.332\up{2.6}$^{*}$\\
Routing using QPP(\rrf{}) & 0.5945\up{33.9}$^{\#}$ & 0.334\up{3.2}$^{\#}$ \\
Weighted QPP Scoring(\wrf{}) & 0.6128\up{38.5} & 0.335\up{3.5} \\

\midrule
\multicolumn{3}{l}{\bf Our approach WITHOUT query rewriting during inference} \\
\midrule
Specialised Ranker (SR) & 0.421\down{5.2}& 0.316\down{2.4}\\
Balanced Score Fusion(\brf{}) & 0.535\up{20.5}$^{*}$ & 0.368\up{13.7}$^{\#}$ \\
Routing using QPP(\rrf{}) & \underline{0.618\up{39.3}}$^{*}$ & \underline{0.382\up{17.9}}$^{*}$ \\
Weighted QPP Scoring(\wrf{}) & \textbf{0.659\up{48.4}}$^{\#}$ & \textbf{0.389\up{20.2}}$^{*}$ \\

\bottomrule        
\end{tabular}

    \caption{\small{Comparison between Baselines and our approaches (\brf{}, \rrf{}, \wrf{}) on \mshard{}. We show the relative improvement of approaches against a baseline (BR) in parentheses. Statistically significant improvements at a level of $95\%$ and $90\%$ are indicated by $*$ and $\#$ respectively~\cite{paired_significance_test}. $^{\dag}$ indicate values taken from \mshard{} leaderboard. The best results are in \textbf{bold} and second is \underline{underlined}.}}
    \vspace{-10mm}
    \label{tab:reranking-msmarco-hard-doc}
\end{table}
\vspace{-2mm}

\section{Results}
\label{sec:results}

\subsection{Effect of Specialized Ranker (SR)}
Table \ref{tab:reranking-msmarco-hard-doc}, shows the comparison of our approaches with baseline models described in Section~\ref{models} and~\ref{ranking_models}. Other baseline model performances are taken from the \mshard{} leaderboard. We also show the performance of our approach with both the original and the rewritten test queries. 
To answer \textbf{RQ1}, from Table \ref{tab:reranking-msmarco-hard-doc} it is clear that just using SR does not outperform the baseline (BR). But when used in conjunction with BR using \brf{}, we see improvement of about \textbf{14\%} and \textbf{21\%} on nDCG@10 and RR respectively. This is because BR does better on easy queries and SR on hard queries as evident with the performance of \rrf{}. 
Additionally, the performance of SR on hard queries can be attributed to its ability to learn relevance patterns specific to hard queries due to its training on contextually enriched queries. 
Further support is evident as we observe a performance improvement from \brf{} to \rrf{}, where in \rrf{} only hard queries are accessed by SR.
While \brf{} has impressive gains, it still does not beat the performance of some baseline models (Electra-MaxP, RM3+Electra-MaxP). 
All our models outperform existing query expansion approaches like \qd{} and \ct{}. On further analysis, we observe that the drop in performance for \qd{} is due to the topic drift in the rewrites obtained from \qd{}.
We observed the same in our experiments and mitigated topic drift by using a passage selector for documents (Section~\ref{method:attention_linear}).

\subsubsection{Effect of Query Rewriting during inference}
We also try inference using rewritten queries, BR infers on original queries and SR on rewritten queries to keep it consistent with their respective training techniques. In Table \ref{tab:reranking-msmarco-hard-doc}, we see that rewriting test queries perform better than BR but do not outperform some other baseline models or our approach using original test queries. This is surprising enough but expected as rewriting hard queries does not always do well~\cite{prf_effectiveness}. We theorize it does badly as the expanded/rewritten query does not have context resulting in topic drift, contrary to the queries SR is trained on which have context.

\mpara{Insight 1}: \textit{Our specialized ranker improves ranking performance on hard queries and also for general queries when used with BR.
}


\subsection{Effectiveness of document ranking using Query Performance Predictor (QPP)}


In Table \ref{tab:reranking-msmarco-hard-doc}, we observe that \brf{} offers lesser gains compared to other approaches. This is primarily because \brf{} does not consider the nature of the query, and it applies equal weights to scores from BR and SR. However, \rrf{} which considers the characteristics of individual queries and perform query routing accordingly provides significant gains. We observe an overall performance improvement of about \textbf{18\%} and \textbf{39\%} on nDCG@10 and RR, respectively. Hence proving that using QPP along with SR and BR for ranking improves performance, answering \textbf{RQ2}.
On closer inspection, we observe that SR provides a significant gain in query performance for hard queries relative to the BR. For instance, the average nDCG@10 for hard queries identified by QPP using BR is 0.278 and on the other hand, SR is 0.345 (\textbf{+24\%} over BR).

\mpara{Insight 2}: \textit{We find that query routing using QPP-based hardness estimation improves ranking performance by \textbf{18\%} on nDCG@10.}

\subsection{Effectiveness of QPP based scoring approach}
\label{subsec:qpp_rf}

To answer \textbf{RQ3}, we evaluate \wrf{} using \mshard{}.
The routing of the queries helps in choosing the ranker based on the nature of the queries. However, since the threshold to determine hardness decides the nature of the query, it could sometimes result in a misclassification of the queries. Using \wrf{}, we can leverage the estimated score directly to weigh the relevance scores from the ranking models that help combine the capabilities of both models for each query. We observe that our proposed approach \wrf{} provides the best performance with a gain of \textbf{20.2\%} in nDCG@10 compared to BR as shown in Table \ref{tab:reranking-msmarco-hard-doc}.  This method even outperforms the SOTA Electra-MaxP~\cite{clark2020electra}. It is interesting to note that the BERT model we use are \bert{}-base models whereas in other baselines BERT-Large and T5 models are more than double the parameters of our model. This improvement of our approach over larger models also proves the effectiveness of QPP score based method(\wrf{}).

\mpara{Insight 3}: \textit{Weighted QPP score-based method combines the best capabilities of both rankers, weighing them according to the hardness of the query and beating SOTA model with \textbf{20\%} improvement.}
\begin{table}[hbt!]
    \centering
    \begin{tabular}{lll}
        \toprule
                & \multicolumn{2}{l}{\textbf{\trecdl{}-Document}} \\
             \midrule
            \textit{Ranking Models}
             &  $\textbf{2019}$ & $\textbf{2020}$ \\
            \midrule

\multicolumn{3}{l}{\bf Baselines} \\
\bert{}-base (BR)& 0.616 & 0.600  \\
\qd{} \cite{wang2023query2doc} & 0.613\up{1.2} & 0.606\up{1}  \\
\ct{} \cite{jagerman2023query} & 0.590\down{4.3}$^{*}$ & 0.599\up{0}  \\
\midrule

\multicolumn{3}{l}{\bf Our Approach} \\
Balanced Score Fusion(\brf{})& 0.622\up{1}$^{\#}$ & \textbf{0.618\up{3.1}}$^{\#}$  \\
Routing using QPP(\rrf{})& 0.622\up{1.1}$^{\#}$ & 0.610\up{1.7}$^{\#}$  \\
Weighted QPP Scoring(\wrf{}) & \textbf{0.625\up{1.5}}$^{*}$ & 0.590\down{1.7}$^{*}$  \\

        \bottomrule        
    \end{tabular}
    \caption{Comparison between Baseline (BR) and our approaches (\brf{}, \rrf{}, \wrf{}) on \trecdl{}-19 and \trecdl{}-20, Document sets. Relative improvement of approaches against a baseline (BR) in parentheses. Statistically significant improvements at a level of $95\%$ and $90\%$ are indicated by $*$ and $\#$ respectively~\cite{paired_significance_test}. The best results are in \textbf{bold}.}
    \vspace{-5mm}
    \label{tab:rank_fusion_msmarco}
\end{table}

\subsection{Effectiveness on general queries}
We also evaluate general ranking datasets like \trecdl{} 19 and \trecdl{} 20, document collections using our approaches. In Table~\ref{tab:rank_fusion_msmarco} we observe a slight performance improvement of our model against baseline, this is because around 5\% of queries in the dataset is hard, hence the contribution of SR in \rrf{} or \wrf{} is very small to make a big difference in the overall ranking. We see \brf{} is the best approach. This illustrates that our approaches work on both general and hard-ranking datasets. We also experiment with rewritten queries and see similar results as in the case of \mshard{}.

We did additional experiments on \mshard{}, \trecdl{} 19, and \trecdl{} 20 on passage datasets, and the results had similar trends as in their respective document collection approaches. For \mshard{} passage, the best performing model was \wrf{} with a gain of \textbf{10.4\%} and \textbf{28.5\%} over BR on nDCG@10 and RR, respectively. Additionally, our approach showed a gain of \textbf{3.1\%} and \textbf{3.4\%} over respective BR models for \trecdl{} 19 and \trecdl{} 20 respectively. For the sake of space, we avoided showing all the results.

\section{Conclusion}
\label{conclusion}

In this work, we propose a framework for improving the document ranking performance for hard queries without sacrificing the performance on other queries. We accomplish this by firstly training a specialized document ranker on hard queries rewritten through context aware query enrichment using LLMs. We then perform query performance estimation using a neural scoring mechanism. Using the query performance scores as indicators of the degree of hardness of the queries, we propose a principled combination of the relevance scores from the base ranker and the specialized ranker. Through extensive experiments on diverse datasets, we demonstrate that the proposed approach offers performance gains for the ranking tasks on hard queries without sacrificing performance on other queries, even outperforming \textbf{SOTA} model on \mshard{} document test set. In the future, we plan to extend our approach to characterize diverse types of queries and also propose an end to end optimization of the query rewriter and ranker.

\renewcommand*{\bibfont}{\scriptsize}
\bibliographystyle{abbrvnat}
\bibliography{reference}

\end{document}